\documentclass{elsart}

\begin{document}

\newcommand{\R} {\,\mbox{Re}\,}
\newcommand{\Tr} {\,\mbox{Tr}\,}

\begin{frontmatter}

\title{On the probability of ferromagnetic strings in 
antiferromagnetic spin chains. }
\author{Alexander G. Abanov} 
 
\address{
Department of Physics \& Astronomy, 
Stony Brook University, \\ Stony Brook, NY 11794-3800, USA}

\author{Vladimir E. Korepin}
\address{C.N. Yang Institute for Theoretical Physics,
Stony Brook University, 
\\ Stony Brook, NY 11794-3840, USA}

\begin{abstract}
We study the probability of formation of ferromagnetic string in the 
antiferromagnetic spin-1/2  XXZ chain. We show that in the limit of 
long strings with weak magnetization per site the  
bosonization technique can be used to address the problem. At zero 
temperature the obtained probability is Gaussian as a function of the  
length of the string. At finite but low temperature there is a 
crossover from the Gaussian behavior at intermediate lengths of 
strings to the  exponential decay for very long strings. 
Although the weak magnetization per site is a necessary small parameter 
justifying our results, the extrapolation of obtained results to the 
case of maximally ferromagnetic strings is in qualitative agreement 
with known numerics and exact results. The effect of an external magnetic field on the probability of formation of ferromagnetic strings is also studied.
\end{abstract}

\begin{keyword}
integrable models, spin chains, correlation functions, bosonization, 
emptiness formation probability, asymptotic behavior
\end{keyword}
\end{frontmatter}


\section{Introduction}
    \label{intro}

Antiferromagnetic spin-1/2  XXZ chain is the first integrable model 
known to be solvable by means of Bethe Ansatz \cite{Bethe-1931,Yang}.
It is described by the Hamiltonian 
\begin{equation}
    H  =J\sum_{j}\left[
    \sigma_{j}^{x}\sigma_{j+1}^{x}+\sigma_{j}^{y}\sigma_{j+1}^{y}
    +\Delta \sigma_{j}^{z}\sigma_{j+1}^{z}\right],
    \label{eq:XXZ}
\end{equation}
where $J>0$ is an exchange integral,  $\Delta$ is an exchange anisotropy, and summation is taken over lattice sites of an infinite spin chain.  In our notations the particular values $\Delta=+1,0,-1$ correspond to isotropic antiferromagnet, XY magnet, and isotropic ferromagnet respectively. It is well known \cite{KBI-1993,Taka,Affleck-1988} that for $-1< \Delta < 1$ (so-called critical regime) the spectrum of excitations around the ground state is gapless and spin-spin correlation functions decay with the distance as a power law. In this paper we consider only  the critical regime $-1< \Delta < 1$.  

Although thermodynamic properties of such a spin chain are well 
understood \cite{Taka}, the exact calculation of correlation functions  
is a long-standing hard problem (see \cite{KBI-1993} and references 
therein). Correlation functions in the  model (\ref{eq:XXZ}) (as well as in 
other integrable models) in thermodynamic limit can be represented as 
combinations of minors and determinant of Fredholm operator of a
special form [integrable integral operator].
From the point of view of this  determinant representation 
the simplest and the most fundamental object is not  a correlation 
function of local fields which is given by minors of Fredholm 
operator, but rather a probability $P_n$ of spontaneous formation of 
magnetization on some space interval (ferromagnetic string)  of the 
length $n$  which is given by 
the determinant of that operator. 
This probability was first introduced in \cite{KIEU-1994} (see also \cite{KBI-1993}) as ``emptiness formation probability" or ``probability of formation of ferromagnetic string" (PFFS). It is given by
\begin{equation}
    P_{n} = \frac{1}{Z}
    \Tr \left\{e^{-\beta H}\prod_{j=1}^{n}\frac{1+\sigma_{j}^{z}}{2}
    \right\},
\end{equation}
where $\beta=1/T$ is an inverse temperature and $Z = \Tr \left\{e^{-\beta H}\right\}$ is a partition function of a spin chain (\ref{eq:XXZ}).

Another approach to correlation functions was developed in 
Kyoto\cite{JM-1995}.
In this approach correlations were represented as multiple integrals. 
Analysis of these integrals also shows that PFFS is the simplest correlation function.

Unfortunately, both determinant and multiple integral representations 
of correlation functions grow extremely complex with the increase of 
the length of a ferromagnetic string. So far the PFFS was calculated exactly 
by means of multiple integral 
representation only up to the length 
$n=5$\cite{BKNS-2002}. Numeric 
results known for this probability suggest\cite{BKNS-2002} that the 
asymptotic behavior is Gaussian $P_n\sim e^{-\alpha n^2}$ at zero 
temperature.
At the moment there are only two exact results known about the large $n$ behavior of $P_n$. In both cases the Gaussian behavior is obtained for $P_n$ at large $n$.
The first case is an  XY spin chain ($\Delta=0$) which is equivalent to free lattice fermions. 
In this case an alternative determinant representation for $P_n$ is available which allows 
to calculate PFFS exactly  for any length of the string\cite{STN-2001}. The large $n$ asymptote of PFFS is Gaussian (see  \cite{STN-2001}) $P_n\sim n^{-1/4} (\sqrt{2})^{-n^2}$.
Another important example is XXZ spin chain at $\Delta =1/2$ . 
In this case PFFS is also known explicitly \cite{Stro} (although it is still a conjecture). In this case the large $n$ behavior is again Gaussian  $P_n\sim n^{-5/36} (\sqrt{3}/2)^{3n^2}$.

On the other hand there is a relatively simple and  very powerful method of 
calculation of asymptote of correlation functions for XXZ spin chains. It is a combination 
of bosonization technique and Bethe Ansatz (for review and references see 
\cite{Affleck-1988,GNT-1998,Luk}).
Namely,   at large distances (in the continuum limit) the XXZ spin chain is 
equivalent to a free boson theory with parameters (e.g., 
compactification radius of bosonic field) which can be determined 
exactly from Bethe Ansatz results for thermodynamics of the chain. 
This approach was successfully used to calculate the long distance 
asymptotes of  local spin-spin correlation functions in spin 
chains \cite{Affleck-1988,GNT-1998,Luk}.

In this paper we apply the bosonization technique to study PFFS 
on the background of  an antiferromagnetic spin chain in the limit of the large length $n$
of the string. 

First, we give a qualitative argument in favor of the Gaussian decay 
of PFFS at zero temperature. We are looking for optimal 
space-time configuration of spins (instanton) which costs the least 
action but gives the maximal magnetization of the string of the 
length $n$ at some moment of time. We use two assumptions. First, we 
assume that the effective theory describing spin chain is Lorentz 
invariant at large distances (it is actually anisotropic but scaling 
of space and time is the same). Second, the extra cost in action of 
having ferromagnetic state is constant per unit space-time area. This 
is a consequence of the first assumption and the fact that the cost 
of energy is constant given by exchange integral per lattice spacing. 
We conclude from these two assumptions that the optimal instanton has 
a shape of a circular (elliptic) droplet of the area $\sim n^2$ with 
constant action cost per unit area. This immediately gives the minimal 
action $\sim n^2$ and  $P_n\sim e^{-\alpha n^2}$ at zero temperature. The fluctuations around the found optimal configuration can only change pre-exponential factor but not the value of exponent itself. At finite 
temperature we expect similar behavior when the length of the string 
is much smaller than an inverse temperature (properly rescaled to 
spatial units). On the other hand, in the limit of long strings one 
of $n$'s in the action of instanton is replaced by the inverse 
temperature $\beta$ (``time" is finite and instanton has a size 
$n\times\beta$. This will produce $P_n\sim e^{-\gamma n}$ -- well 
known result\cite{KBI-1993} with $\gamma$ proportional to the density 
of free energy of an antiferromagnetic state.

Unfortunately, bosonization is not directly applicable to finding PFFS --- the probability of 
formation of maximally ferromagnetic string. 
The reason is that the maximal magnetization corresponds to the large 
(compared to an inverse lattice spacing) gradient of the bosonic field 
which violates the continuum approximation. It turns out, however, 
that one can use this technique to find the probability of formation 
of weakly ferromagnetic strings. For weakly ferromagnetic strings our 
calculations justify the qualitative argument in favor of Gaussian decay 
given above and the obtained results 
are in qualitative agreement with numerics \cite{BKNS-2002,NS-num} 
and  exact results \cite{STN-2001,Stro}  for the strings with maximal magnetization.

In Sec. \ref{sec:Bos} we briefly review the well known results of bosonization of XXZ spin chain. In Sec. \ref{sec:Cal} we present our calculation of the probability of formation of weakly ferromagnetic string. We summarize the results of this calculation in Sec. \ref{sec:Res} and compare them with available numeric and exact results in Sec. \ref{sec:Dis}. We relegated  the calculation in the presence of an external magnetic field and the calculation for strings with uniform magnetization to Appendices \ref{app:exres}, \ref{app:um}.

\section{Bosonization}
    \label{sec:Bos}

Bosonization of the continuum limit of XXZ chain gives the 
action (for review and references see \cite{Affleck-1988,GNT-1998})
\begin{equation}
    S = \int_{-\infty}^{+\infty}dx\int_{0}^{\beta}d\tau\,
    \left[\frac{1}{2v}(\partial_{\tau}\phi)^{2} 
    +\frac{v}{2}(\partial_{x}\phi)^{2}  \right],
 \label{eq:bosactraw}
\end{equation}
where $x=j$ is a continuous coordinate along the chain, and lattice spacing is taken as a unity
$a=1$, $\tau$ is an 
imaginary time ($\hbar=1$), $v$ is a spin wave velocity, 
and $R$ is a radius of 
compactification of free Bose field $\phi$: $\phi\equiv \phi +2\pi 
R$. The parameters $R$ and $v$ of the model (\ref{eq:bosactraw})  are known exactly from the Bethe ansatz solution of the model (\ref{eq:XXZ}) \cite{KBI-1993}
\begin{eqnarray}
    2\pi R^{2} &=& 1 
    -\frac{1}{\pi}\cos^{-1}\Delta,
 \label{eq:Rexact} \\
    v &=& 2\pi\frac{\sqrt{1-\Delta^2}}{\cos^{-1}\Delta}J.
 \label{eq:vsexact}
\end{eqnarray}
In the following we put\footnote{One can always restore 
the dependence on spin wave velocity replacing at the end of calculation $\beta\to 
\beta\ v$.} $v=1$.
\begin{equation}
    S = \int_{-\infty}^{+\infty}dx\int_{0}^{\beta}d\tau\,
    \frac{1}{2}(\partial_{\mu}\phi)^{2} .
 \label{eq:bosact}
\end{equation}
The original spin operators are given in terms of bosonic field as
\begin{eqnarray}
    \sigma_{j}^{z} &=& \frac{1}{\pi R}
    \frac{\partial\phi}{\partial x}
    +const\cdot (-1)^{j} \cos\frac{\phi}{R}
 \nonumber \\
    \sigma_{j}^{\pm} &=& e^{\mp i2\pi R\tilde\phi}\left[const \cdot
    \cos\frac{\phi}{R} + C(-1)^{j}\right],
 \nonumber
\end{eqnarray}
where 
\begin{eqnarray}
    \phi(\tau,x)&=&\phi_{L}(z)+\phi_{R}(\bar{z})
 \nonumber \\
    \tilde\phi(\tau,x)&=&\phi_{L}(z)-\phi_{R}(\bar{z})
 \nonumber
\end{eqnarray}
and $z=x+i\tau$, $\bar{z}=x-i\tau$ are complex coordinates.

Let us consider a long segment of the spin chain of the length $n\gg 1$. 
The total magnetization of the segment at some time $\tau$ 
is given by $m_n(\tau)=\sum_{j}\sigma_{j}^{z}
=\frac{1}{\pi R}\int dx\, \partial_{x}\phi =\frac{1}{\pi 
R}\left[\phi(n,\tau)-\phi(0,\tau)\right].$ The maximal magnetization 
(ferromagnetic string) corresponds to $\sum_{j}\sigma_{j}^{z}=n$. We are 
looking for the probability of spontaneous formation of the string of spins of the length $n$ 
and average magnetization per site $c$ ($c=1$ corresponds to the 
maximal magnetization $\sigma^{z}=1$ per site). We require therefore,
\begin{equation}
    \left[\Delta\phi\right]\equiv \phi(n)-\phi(0) = c n\pi R.
 \label{eq:bch0}
\end{equation}
Here we denoted the change of the bosonic field as 
$\left[\Delta\phi\right]$ to avoid a confusion with the Laplacian of 
the same field $\Delta\phi$. Notice that the typical value of the 
gradient of $\phi$ along the string is $\partial_{x}\phi \sim c\pi R$. 

Bosonization is an essentially continuum approach which is valid only 
in the limit when $\partial_{x}\phi\sim \partial_{\tau}\phi \ll \pi R$. 
This requirement gives the condition $c\ll 1$ with the original 
problem of finding the probability of maximally ferromagnetic string 
to be beyond the range of applicability of bosonization approach. In 
the following we assume $c\ll 1 $ and look for the probability of 
{\em weakly ferromagnetic strings}.  The following remark is in place. In contrast to the problem of maximally ferromagnetic string the problem of finding the probability of a weakly ferromagnetic string is not yet completely defined. Namely, one can search for the probability specifying the total magnetization of the string only (\ref{eq:bch0}) or one can fix the magnetization profile along the string $\sigma^z(x) = c\pi R f(x)$ with arbitrary real function $f(x)$ satisfying $f(0)=0$ and $f(n)=n$. In the following we adopt the first interpretation (obviously it is equivalent to solving the problem with given $f(x)$ and then averaging over all $f(x)$). The problem in the second interpretation can also be straightforwardly solved by presented methods. We consider the particularly interesting case of uniform magnetization in Appendix \ref{app:um}.
To use the bosonization method one should also require $n\gg 1$, i.e., long strings. Finally the free Bose field approximation can be used only at large space-time 
scales where irrelevant (and marginal irrelevant for XXX case) 
operators coming from $\sigma^{z}\sigma^{z}$ term renormalize to zero. 
It turns out, however, that this happens at scales of the order of 1 
and imposes no additional restriction.

\section{Calculation}
    \label{sec:Cal}
                          
The probability to have a ferromagnetic string is given by 
\begin{equation}
 \label{eq:PnOr}
    P_{n} = \frac{1}{Z}\int D\phi\, 
    \delta(\left[\Delta\phi\right] -c n\pi R) e^{-S[\phi]},
\end{equation}
where $S[\phi]$ is given by  
(\ref{eq:bosact})  and $Z=\int D\phi\,  e^{-S[\phi]}$. We 
represent delta function as 
\begin{equation}
 \label{eq:PnLagr}
    P_{n} = \frac{1}{Z}\int_{-\infty}^{+\infty}\frac{d\lambda}{2\pi}
   \int D\phi\, 
    e^{-S[\phi]-i\lambda(\left[\Delta\phi\right]-c n\pi R)}.
\end{equation}

The obtained integral in $\phi$ is Gaussian and is determined 
exactly by its saddle point which can be obtained from
\begin{equation}
    \Delta\phi_0 = i\lambda\left[\delta^2(z-n)-\delta^2(z)\right]
 \label{eq:lapleq}
\end{equation}
with periodic boundary conditions in time
\begin{equation} 
    \phi(z+i\beta) =\phi(z) +2\pi R w,
 \label{eq:bcpertop}
\end{equation}
where $w$ is any integer number (winding number). The expression $\delta^2(z-n) = \delta(x-n)\delta(\tau)$ denotes two-dimensional Dirac delta function.
In this paper we consider only the limit of an
infinitely long spin chain. Then, only topologically trivial $w=0$ 
sector will contribute to the probability because the action for the sectors with 
$w\neq 0$ is proportional to the length of the chain. Therefore, we require
\begin{equation} 
    \phi(z+i\beta) =\phi(z).
 \label{eq:bcper}
\end{equation}

\subsection{Zero temperature}

Let us first consider the case of zero temperature $\beta\to \infty$. 
Then (\ref{eq:bcper}) is irrelevant (instead we require the decay of $\partial _\mu \phi$ at infinity) and the solution of (\ref{eq:lapleq}) is given by
\begin{equation}
    \phi_0(z) = \frac{i\lambda}{2\pi} \R \ln\frac{z-n}{z}
 \label{eq:phi0}
\end{equation}
and being substituted into (\ref{eq:PnLagr}) gives  the 
Green's function of free boson at zero temperature
$\langle e^{-i\lambda\phi(n)}e^{i\lambda \phi(0)}\rangle$ or 
\begin{equation}
 \label{eq:Pnalpha}
    P_{n} = \int_{-\infty}^{+\infty}\frac{d\lambda}{2\pi}
    e^{-\frac{\lambda^2}{2\pi}\ln\frac{n}{r_0}+i\lambda  \frac{2n}{Q}},
\end{equation}
where we introduced new notation
\begin{equation}
    Q = \frac{2}{c\pi R} \gg 1
  \label{eq:Q}
\end{equation}
and  $r_0$ is some short distance cutoff.
This integral is in turn determined by its saddle point
\begin{equation}
    -\frac{i\lambda_0}{2\pi} = \frac{1}{ Q\ln\frac{n}{r_0}}
 \label{eq:l0}
\end{equation}
and is given by
\begin{equation}
 \label{eq:Pnr0}
    P_{n} = \frac{1}{2\ln\frac{n}{r_0}} 
    e^{-\frac{2\pi n^2}{Q^2 \ln\frac{n}{r_0}}}.
\end{equation}
Let us now find a short distance cutoff $r_{0}$. On one hand it must 
be at least bigger than $1$ -- the lattice scale. On the other hand we 
can trust our bosonization approach only if $\partial_\mu\phi\ll \pi R$. From  
(\ref{eq:phi0}) we have on saddle point configuration
$$ (\partial_{\mu}\phi_{0})^{2} = \left|
\frac{i\lambda_{0}}{2\pi}\left(\frac{1}{z-n}-\frac{1}{z}\right)
\right|^{2}.$$
This gradient diverges near the ends of the string. At points close 
to $z=0$ (similar for points close to $z=n$) we have
\begin{equation}
    (\partial_{\mu}\phi_{0})^{2} = \left|
    \frac{i\lambda_{0}}{2\pi}\right|^{2}
    \frac{1}{r^{2}},
\end{equation}
where $r=|z|$ (or $r=|z-n|$ near the other end of the string).
We obtain $r_{0}$ from
\begin{equation}
    |\partial_{\mu}\phi_{0}|_{r=r_{0}} = -\frac{i\lambda_{0}}{2\pi}
    \frac{1}{r_{0}} = \pi R.
\end{equation}
Using (\ref{eq:l0}) we obtain
\begin{equation}
    -\frac{i\lambda_0}{2\pi}=\pi R r_0=\frac{n }{ Q } \ln\frac{n}{r_0}.
\end{equation}
We solve the latter equation by iterations and obtain
\begin{equation}
    r_{0}=\frac{cn}{2\ln c^{-1}}
    \left(1+O\left(\frac{1}{\ln c^{-1}}\right)\right). 
 \label{eq:r0}
\end{equation}
In (\ref{eq:r0}) and in the following we have actually replaced $O\left(\frac{\ln\ln c^{-1}}{\ln c^{-1}}\right)$  by $O\left(\frac{1}{\ln c^{-1}}\right)$. 
Equation (\ref{eq:r0}) gives the short distance cutoff for long strings $n\gg c^{-1}\ln c^{-1}$. If strings are not so long one should use the lattice spacing cutoff $r_0\sim 1$. We have
$$
    r_{0}=\mbox{max}\left\{1,\frac{cn}{2\ln c^{-1}}\right\}. 
$$
In the following we consider only sufficiently long strings $n\gg c^{-1}\ln c^{-1}$ and use (\ref{eq:r0}).
We find therefore, that bosonization 
approach is valid everywhere except for the vicinity of the ends of 
the string which is defined by the radius $r_{0}$ from (\ref{eq:r0}) ($r_0\ll n$). 
Substituting the found value of $r_{0}$ into (\ref{eq:Pnr0}) we obtain (we keep only exponential dependence)
\begin{eqnarray}
     P_{n} &\sim & e^{-\alpha n^{2}}, \;\mbox{when}\; n\gg c^{-1}\ln c^{-1}
 \label{eq:Pn} \\
    \alpha &=& \frac{2\pi}{Q^{2}\ln c^{-1}}
 \label{eq:alpha}
\end{eqnarray}

Let us now estimate the corrections to our results coming from the vicinities (of the radius $r_0$) of the ends of the string where bosonization approach is not applicable. First of all, we find that the optimal configuration $\phi_0$ is given by (\ref{eq:phi0}), (\ref{eq:l0}), and (\ref{eq:r0}) as   
\begin{equation}
    \phi_0(z) = \frac{n}{Q\ln c^{-1}} \R \ln \frac{z}{z-n}.
 \label{eq:phiopt}
\end{equation}
We obtain
$$
[\Delta\phi_0]\approx \phi_0(n-r_0)-\phi_0(r_0) \approx 2\frac{n}{Q} 
$$
which satisfies (as expected) the boundary conditions (\ref{eq:bch0}) with the accuracy  $1/\ln c^{-1}$. We estimate an error coming from the vicinity to the end of the string assuming the maximal gradient $\partial_x\phi_0\sim \pi R$ in that vicinity $\phi_0(r_0)-\phi_0(0)  \sim r_0  \partial_x \phi_0 \sim \pi Rr_0  \sim n/(Q\ln c^{-1})\sim \left[\Delta\phi\right] \frac{1}{\ln c^{-1}}$ with $\left[\Delta\phi\right]$ from (\ref{eq:bch0}). We obtain that in the phase difference calculations one can neglect the vicinities of the strings with our usual accuracy $\sim \frac{1}{\ln c^{-1}}$. Secondly, let us estimate the contribution to the action coming from those vicinities. Again we use $\partial_\mu \phi_0 \sim \pi R$ and find correction to the action $ r_0^2 (\partial_\mu\phi_0)^2\sim (\pi R r_0)^2 \sim \frac{ n^2}{Q^2\ln^2 c^{-1}} \sim \alpha n^2 \frac{1}{\ln c^{-1}}$. We conclude that with the accuracy $\frac{1}{\ln c^{-1}}$ one can neglect the regions where bosonization breaks down. 

Therefore, in the case of zero temperature the probability to have a weakly ferromagnetic string of the length $n\gg c^{-1}\ln c^{-1}\gg1$ in the antiferromagnetic XXZ spin chain is Gaussian  (\ref{eq:Pn}), (\ref{eq:alpha}) and the  small parameter justifying our bosonization calculation is $1/\ln c^{-1}$.

\subsection{Finite temperature}
  \label{temperature}

In this section we consider small but finite temperature $T$ such that $\beta \gg 1$ 
(in full units $\beta v\gg 1$ and one can still use continuum 
limit and bosonization technique. The only thing which is modified in 
this case compared to the case of zero temperature is that one should 
solve the same problem (\ref{eq:PnLagr}) on a cylinder $0<\tau<\beta$ with periodic 
boundary conditions (\ref{eq:bcper}).   
    
Similarly to the case or zero temperature we obtain the equation (\ref{eq:lapleq}) for the saddle point configuration. With boundary conditions (\ref{eq:bcper}) its solution  is given by
\begin{equation}
    \phi_0(z) = \frac{i\lambda}{2\pi} \R \ln\frac{\sinh\frac{\pi}{\beta}(z-n)}{\sinh\frac{\pi}{\beta}z}.
 \label{eq:phiT}
\end{equation}
Notice that  in the limit $n\ll \beta$ the main contribution to 
(\ref{eq:PnLagr}) comes from distances $|z|\sim n\ll \beta$ and one can expand hyperbolic sines in  (\ref{eq:phiT}) obtaining (\ref{eq:phi0}). Thus, in the limit of low temperatures we restore the results obtained for zero temperature. 
For the probability of the string (compare to eq.(\ref{eq:Pnalpha})) we obtain
\begin{equation}
 \label{eq:PnalphaT}
    P_{n} = \int_{-\infty}^{+\infty}\frac{d\lambda}{2\pi}
    e^{-\frac{\lambda^2}{2\pi}\ln\frac{\sinh(\pi n/\beta)}{\pi r_0/\beta}+i\lambda  \frac{2n}{Q}},
\end{equation}
where we again assumed $r_0\ll \beta$ in treating short distance cutoff. One can easily recognize in the first exponent the finite temperature correlator of free Bose fields $\langle e^{-i\lambda\phi(n)}e^{i\lambda\phi(0)} \rangle$.
Integral over $\lambda$ is determined by the saddle point
\begin{equation} 
     -\frac{i\lambda_0}{2\pi} 
    =\frac{n}{Q\ln\frac{\sinh(\pi n/\beta)}{\pi r_0/\beta}}
 \label{eq:l0T}
\end{equation}
and we find (similarly to eq.(\ref{eq:Pnr0}))
\begin{equation}
 \label{eq:PnrT}
    P_{n} = \frac{1}{2\ln\frac{\sinh(\pi n/\beta)}{\pi r_0/\beta}} 
    e^{-\frac{2 \pi n^2}{Q^2 \ln\frac{\sinh(\pi n/\beta)}{\pi r_0/\beta}}}.
\end{equation}
The gradient of the field
\begin{equation} 
 \label{eq:gradT}
     \partial_x \phi_0-i\partial_\tau\phi_0 = \frac{i\lambda}{2\beta}\left[
    \coth\frac{\pi}{\beta}(z-n)
    -\coth\frac{\pi}{\beta}z
    \right].
\end{equation}
diverges near the ends of the string and  we again introduce the short distance cutoff $r_0$ as  the scale such that $\partial_x\phi_0(r_0)=-\frac{i\lambda}{2\beta}
\coth\frac{\pi}{\beta}r_0 =\pi R$. In the following we show that $r_0\ll \beta$. Using this inequality we obtain $\partial_x\phi_0(r_0)=-\frac{i\lambda}{2\pi r_0}=\pi R$ and  find 
\begin{equation}
   \pi R r_0=-\frac{i\lambda}{2\pi}.
 \label{eq:r0T}
\end{equation} 

Short distance cutoff $r_0$ is defined by (\ref{eq:r0T}) and (\ref{eq:l0T}) as
\begin{equation} 
     \pi R r_0 =\frac{n}{Q\ln\frac{\sinh(\pi n/\beta)}{\pi r_0/\beta}}.
 \label{eq:r0rec}
\end{equation}
In the following we are interested in sufficiently long strings $n\gg c^{-1}\ln c^{-1}$ and sufficiently low temperatures $\beta\gg c^{-1}\ln c^{-1}$. It is easy to show that in these limits the solution of (\ref{eq:r0rec}) satisfies $r_0\gg 1 $.
Using (\ref{eq:r0rec}) we rewrite (\ref{eq:PnrT}) as
\begin{equation}
 \label{eq:PnrTsimple}
    P_{n} = \frac{\pi R r_0 Q}{2n} 
    e^{-\frac{2 \pi (\pi Rr_0) n}{Q}}.
\end{equation}
Equations (\ref{eq:PnrTsimple}) and (\ref{eq:r0rec}) give the probability of weak ferromagnetic string at finite temperature.

We solve (\ref{eq:r0rec}) by iterations and obtain
\begin{equation} 
     \pi R r_0 =\frac{n}{Q\ln\left(\pi R Q\frac{\sinh(\pi n/\beta)}{\pi n/\beta}\right)}.
 \label{eq:r0sol}
\end{equation}
Next iterations give corrections which are small by parameter 
$$\left[\ln\left(\pi R Q\frac{\sinh(\pi n/\beta)}{\pi n/\beta}\right)\right]^{-1}< 
\left[\ln c^{-1}\right]^{-1} \ll 1.$$
Using (\ref{eq:r0sol}) we obtain the leading exponent from (\ref{eq:PnrTsimple})
\begin{equation}
 \label{eq:PnrTsol}
    P_{n} \sim  e^{-\frac{2 \pi  n^2}{Q^2 \ln \left(\pi R Q\frac{\sinh(\pi n/\beta)}{\pi n/\beta}\right)}}.
\end{equation}
 Let us now check that  the use of bosonization to derive (\ref{eq:PnrTsol}) is justified.
Firstly, for bosonization to be applicable the main contribution to the magnetization should come from the scales large than $r_0$. Namely, using (\ref{eq:phiT}), (\ref{eq:l0T}), and (\ref{eq:r0rec}) we obtain
\begin{eqnarray} 
     [\Delta\phi_0]\approx \phi_0(n-r_0)-\phi_0(r_0) &=& \frac{2n}{Q}
 \label{eq:pd}
\end{eqnarray}
and magnetization (change of $\phi_0$) coming from the scale larger than $r_0$ and given by (\ref{eq:pd}) is indeed the total magnetization of the string. On the other hand the contribution from segments $[0,r_0]$ and $[n-r_0,n]$ to the total magnetization is of the order of $\pi R r_0$ (it is determined by maximal gradient equal to $r_0\partial_x\phi_0\sim \pi R r_0$) and we obtain from (\ref{eq:pd}) the required consistency condition
$$\pi R r_0 \ll  \frac{n}{Q}$$
which together with already used assumption $r_0\ll \beta$ gives
\begin{equation}
    r_0\ll \mbox{min}(cn,\beta).
 \label{eq:cc}
\end{equation}
Secondly, calculating the optimal action (the value in the exponent of (\ref{eq:PnrTsimple})) we neglected contributions from the $r_0$-vicinities of the ends of the string. We estimate that neglected action as $r_0^2(\partial_\mu\phi)^2\sim (\pi R r_0)^2$ with maximal gradient of the order of $\pi R$.
It is easy to see that if this consistency condition is satisfied one can  neglect this contribution as it is smaller than the action coming from larger distances
$(\pi R r_0)^2\ll n(\pi R r_0)/Q$.

We conclude that the use of bosonization is legitimate if the consistency condition (\ref{eq:cc}) is satisfied. It is easy to check that the solution (\ref{eq:r0sol}) indeed satisfies this condition in the limit of sufficiently long $n\gg c^{-1}\ln c^{-1}$ strings and sufficiently low $\beta\gg c^{-1}\ln c^{-1}$ temperature. Under these conditions the equation (\ref{eq:PnrTsol}) gives the probability of spontaneous formation of a ferromagnetic string of  an arbitrary length  and at arbitrary temperature. Below we consider two important limiting cases. 

\subsubsection{Short   strings: $1\ll n\ll \frac{\beta}{\pi}\ln c^{-1}$}

Let us assume that $\frac{\sinh(\pi n/\beta)}{\pi n/\beta}\ll \pi R Q$. This is equivalent to taking not very long strings $n\ll \frac{\beta}{\pi}\ln c^{-1}$. In this case we drop the combination
$\frac{\sinh(\pi n/\beta)}{\pi n/\beta}$ from (\ref{eq:PnrTsol}) and (\ref{eq:r0sol}) and obtain
\begin{equation} 
    r_0=\frac{cn}{2\ln c^{-1}}
 \label{eq:r0Tshort}
\end{equation}
and
\begin{equation}
 \label{eq:PnrTshort}
    P_{n} \sim  e^{-\frac{2\pi n^2}{Q^2 \ln c^{-1}}}.
\end{equation}
Equations (\ref{eq:r0Tshort}-\ref{eq:PnrTshort}) reproduce the results (\ref{eq:r0}-\ref{eq:alpha}) we obtained in the limit of zero temperature. 

\subsubsection{Very long strings: $n\gg \frac{\beta}{\pi}\ln c^{-1}$}

In the opposite limit of long strings $\frac{\sinh(\pi n/\beta)}{\pi n/\beta}\gg \pi R Q$ (or $n\gg \frac{\beta}{\pi}\ln c^{-1}$) we replace $\ln \left(\pi R Q\frac{\sinh(\pi n/\beta)}{\pi n/\beta}\right)\to \frac{\pi n}{\beta}$ and obtain from (\ref{eq:r0sol},\ref{eq:PnrTsol})
\begin{equation} 
    r_0=\frac{\beta c}{2\pi}
\end{equation}
and 
\begin{equation}
 \label{eq:PnrTlong}
    P_{n} \sim  e^{-\frac{2 \beta n}{ Q^2}}.
\end{equation}
In this limit the probability decays exponentially with the length of the string as it is expected\cite{KBI-1993}.

\section{Results}
    \label{sec:Res}

We summarize the results of our calculation as
\begin{equation}
 \label{eq:Pn_sum}
    P_{n} \sim  e^{-\frac{2 \pi  n^2}{Q^2 \ln \left(c^{-1} \frac{\sinh(\pi n/\beta)}{\pi n/\beta}\right)}}
   \sim\left\{
    \begin{array}{lll}
        e^{-\alpha n^2}, & \;\;\alpha = \frac{2\pi}{Q^2 \ln c^{-1}} ,
       &\;\; n\ll \frac{\beta}{\pi}\ln c^{-1} ,
     \\
        e^{-\gamma n}, & \;\;\gamma = \frac{2 \beta }{ Q^2} ,
       & \;\; n\gg \frac{\beta}{\pi}\ln c^{-1} ,
   \end{array}
    \right.
\end{equation}
where $P_n$ is a probability of finding the weakly ferromagnetic string of the length $n$ and total magnetization $m_n=cn$ in an antiferromagnetic XXZ spin chain. The result is valid for small per site magnetization $c\ll 1$, long  strings $n\gg c^{-1}\ln c^{-1}\gg 1$ and  at finite but sufficiently low temperatures $1/T=\beta \gg c^{-1}\ln c^{-1}\gg 1$ ($\beta$ is measured in units of $1/v$ with $v$ from (\ref{eq:vsexact})). The large parameter $Q=2/(c\pi R) \gg 1$ with compactification radius $R$ from (\ref{eq:Rexact}). In the limit of zero temperature the actual small parameter justifying our calculations is $1/\ln c^{-1}\ll 1$.

It is interesting to notice from  (\ref{eq:Pn_sum}) that the crossover from exponential to Gaussian decay of probability occurs not at the scale $n\sim \beta$ but at much larger (for weakly ferromagnetic strings $c\ll 1$) scale $n\sim \frac{\beta}{\pi}\ln c^{-1}$. There is a simple reason for such a behavior. We fixed only the total magnetization of the string. It turns out that the magnetization is concentrated mostly at the ends of the string, namely, within the distance of $n/\ln c^{-1}$ to the ends of the string. Therefore, the total length of the string is irrelevant and the crossover takes place only when the new scale becomes of the order of $\beta$, i.e.,  $n\sim \beta\ln c^{-1}$.

We also present in this section the results we obtain in Appendix \ref{app:um} (see (\ref{eq:finalpha},\ref{eq:fingamma}) for the probability of formation of a weak ferromagnetic string with uniform magnetization 
\begin{equation}
 \label{eq:Pn_sumum}
    P_{n}  
   \sim\left\{
    \begin{array}{lll}
        e^{-\alpha n^2}, & \;\;\alpha = \frac{\pi}{2Q^2} ,
       &\;\; n\ll \beta ,
     \\
        e^{-\gamma n}, & \;\;\gamma = \frac{2 \beta }{ Q^2}, 
       & \;\; n\gg \beta .
   \end{array}
    \right.
\end{equation}
The small parameter justifying our calculation in the limit of zero temperature is $c^2$ in this case. It is much smaller for  $c\ll 1$ than the parameter $1/\ln c^{-1}$ for the case of free string---the string with only the total magnetization fixed. 

It is easy to generalize our results to the case when external magnetic field $h$ is present. The results are still given by (\ref{eq:Pn_sum})  or (\ref{eq:Pn_sumum}) with the replacement $c\to \tilde{c}$ where $\tilde{c}$ is given in (\ref{eq:ctilde}) (this replacement means that only the deviation from the background magnetization induced by magnetic field is relevant). Also, all other parameters $v,R$ must be taken from the exact solution in non-zero magnetic field (see Appendix \ref{app:exres} for details).

\section{Discussion}
    \label{sec:Dis}

We studied the probability of formation of ferromagnetic string in XXZ spin chains. The method of continuum bosonization we used is not applicable to the case of a maximal magnetization of the string. However, we were able to obtain the behavior of this probability in the limit of a weakly ferromagnetic strings. In this limit we have a controllable approximation with a small parameter related to the small per site magnetization of the string. Our main results are presented in (\ref{eq:Pn_sum},\ref{eq:Pn_sumum}) and in Appendix \ref{app:exres} in the presence of an external magnetic field.

These results: crossover from Gaussian behavior at low temperatures (short strings) to an exponential decay at high temperatures (long strings) are in qualitative agreement with numeric results \cite{BKNS-2002,NS-num} and known exact results \cite{STN-2001,Stro}. We have also presented the qualitative argument in favor of such a behavior in Sec. \ref{intro}.

Let us now see how far off we are quantitatively. We will try to predict the rate of  Gaussian decay $\alpha$ by (illegally) extending our results (\ref{eq:Pn_sumum}) obtained in the limit of a weakly ferromagnetic string $c\ll 1$ with a uniform magnetization to the case of maximally ferromagnetic string $c=1$ (which obviously has a uniform magnetization). We choose (\ref{eq:Pn_sumum}) versus (\ref{eq:Pn_sum})  for such an interpolation because we it to be more precise for not very small $c$ because of a milder divergence of the optimal field configuration in the case of a string with a uniform magnetization (see Appendix \ref{app:um}).

We take $c=1$ (or $Q=2/\pi R$) in (\ref{eq:Pn_sumum}) and obtain
$\alpha =\pi^3 R^2/8$ with the exact value of compactification radius $R$ from (\ref{eq:Rexact}). This result is qualitatively correct giving $\alpha=0$ at $R=0$ which corresponds to the ferromagnetic limit $\Delta=-1$. In this limit we do expect a disappearance of  a Gaussian decay of the probability of PFFS (the range $\Delta<-1$ corresponds to Ising ferromagnetic behavior). However, the exact numeric result is not justified\footnote{We are grateful to S. Lukyanov for stressing this point to us.}.

We  use $\alpha =\pi^3 R^2/8$ with (\ref{eq:Rexact}) and write
\begin{equation}
    \delta \equiv e^\alpha  = e^{\frac{\pi^2}{16} \left(
     1-\frac{1}{\pi}\cos^{-1}\Delta\right)},
 \label{eq:interp}
\end{equation}
where new parameter $\delta$ is introduced so that $P_n\sim \delta^{-n^2}$. In fact, one can expect any coefficient of the order of one instead of $\pi^2/16$ in (\ref{eq:interp}).
In the following table we compare the rates of Gaussian decay of PFFS in XXZ spin chains 
calculated from (\ref{eq:interp}) with two exact results known for $\Delta=0,\frac{1}{2}$ \cite{STN-2001,Stro}. 
\vspace{0.2cm}
\begin{center}
\begin{tabular}{|c|l|c|}
\hline
    \multicolumn{3}{|c|}{\bfseries The rate of Gaussian decay $P_n\sim \delta^{-n^2}$. }
    \\
    \hline
    $ \Delta$ &  $\delta$ (exact) & $\delta$ from Eq.(\ref{eq:interp})  \\ \hline
    \hline
    $0$   &  $1.4142\ldots$ ($=\sqrt{2}$ \cite{STN-2001})  & $1.36$
 \\ \hline
    $1/2$  &   $1.5396\ldots$ ($=\left(\frac{2}{\sqrt{3}}\right)^3$ \cite{Stro})  & $1.51$
 \\ \hline
\end{tabular}
\vspace{0.5cm}
\end{center}

We see from this table that the agreement is reasonably good even quantitatively. 
The comparison of (\ref{eq:interp}) with known numerical results\cite{BKNS-2002,NS-num} obtained for different values of anisotropy $\Delta$ also gives a good agreement  ($\sim 10\%$).

We conclude with the remark that in addition to giving an insight to the nature of a Gaussian decay of the probability of formation of ferromagnetic strings in spin models the problem we considered is interesting on its own. For example, there are known systems where the dopant electron propagation is disfavored  by  the background antiferromagnetic state. In such situations the electron propagation will be assisted by spontaneous formation of ferromagnetic strings. The actual magnetization profile will be determined by the competition between the benefit for the electron propagation and the action cost for the necessary ferromagnetic fluctuation. These are the details of a specific model which will determine the magnetization profile of an optimal fluctuation and a weakly ferromagnetic string is one of the candidates.

\textbf{Acknowledgments}:
We are very grateful to I. Aleiner, S. Lukyanov, A. Tsvelik, O. Starykh, and A. Zamolodchikov for stimulating discussions. We specially thank Y. Nishiyama and M. Shiroishi for sending us their numerical data \cite{NS-num} for the rate of Gaussian decay of PFFS at zero temperature and different values of an anisotropy and S. Lukyanov for sending us the second paper in \cite{Luk} prior to publication. A.G.A. would like to thank the Alfred P. Sloan Foundation for financial support. V.E.K. was supported by NSF PHY-9988566.


\appendix

\section{The effect of an external magnetic field}
\label{app:exres}

In this Appendix we show how to generalize the results obtained in Sec. \ref{sec:Cal} in the presence of an external magnetic field. Magnetic field is introduced into XXZ model as
\begin{equation}
    H  =J\sum_{j=1}^{N}\left[
    \sigma_{j}^{x}\sigma_{j+1}^{x}+\sigma_{j}^{y}\sigma_{j+1}^{y}
    +\Delta \sigma_{j}^{z}\sigma_{j+1}^{z}\right]
    -h\sum_{j=1}^{N}\sigma_{j}^{z}.
    \label{eq:XXZh}
\end{equation}
The bosonization of this model gives\cite{Affleck-1988,GNT-1998}
\begin{equation}
    S = \int_{-\infty}^{+\infty}dx\int_{0}^{\beta}d\tau\,
    \left[\frac{1}{2}(\partial_{\mu}\phi)^{2}  -\frac{h}{\pi 
    R}\partial_{x}\phi \right],
 \label{eq:bosacth}
\end{equation}
where we again use units in which  $v=1$ (at the end of calculation $h\to 
h/v$).

We are going to minimize (\ref{eq:bosacth}) with the condition (\ref{eq:bch0}) on the total magnetization of the string.
We notice that the change of variables 
\begin{equation}
    \phi = \phi'+\frac{h}{\pi R}x.
 \label{eq:htr}
\end{equation} 
transforms (\ref{eq:bosacth}) and (\ref{eq:bch0}) into
(we omit primes and drop  an additive constant)
\begin{eqnarray}
    S &=& \int_{-\infty}^{+\infty}dx\int_{0}^{\beta}d\tau\,
    \frac{1}{2}(\partial_{\mu}\phi)^{2},
 \label{eq:bosacthresc} \\
    \left[\Delta\phi\right] &=& \frac{2n}{Q},
 \label{eq:bchf} 
\end{eqnarray}
where
\begin{eqnarray}
    Q &=& \frac{2}{\tilde{c}\pi R}
 \label{eq:Qtilde} \\
    \tilde{c} &=& c-\frac{h}{\pi^{2}R^{2}}.
 \label{eq:ctilde}
\end{eqnarray}
Comparing (\ref{eq:bosact},\ref{eq:bch0}) and (\ref{eq:bosacthresc}-\ref{eq:ctilde}) we 
notice that the only effect of an 
external magnetic field on the probability of ferromagnetic string (\ref{eq:Pn_sum}) is 
just some effective renormalization of parameter $c$ as well as 
the renormalization of  $R$ and $v$. The renormalization of the parameter $c$ has a very simple physical reason. An external magnetic field results in the uniform background magnetization $h/(\pi^2 R^2)$. The effective renormalization (\ref{eq:ctilde}) means that for the probability of the string it is the deviation of magnetization from the equilibrium one which is  relevant.
The dependence of parameters $v,R$  on a weak\footnote{The weakness of magnetic field $h/\pi^2 R^2 \ll 1$ (or in full units $h/\pi^2 R^2 v\ll 1$) is required so that the field transformation (\ref{eq:htr})  is within the applicability of bosonization approach.} magnetic field is also known\cite{KBI-1993}
from the Bethe Ansatz solution.

\section{Ferromagnetic strings with  uniform magnetization}
 \label{app:um}

The probability of formation of a weak ferromagnetic string is given by the minimum of action (\ref{eq:bosact}) with corresponding boundary conditions on Bose field $\phi(z)$. In Sec. \ref{sec:Cal} we used (\ref{eq:bch0}) as a boundary condition fixing, therefore, only the total magnetization of the string. It is straightforward to minimize (\ref{eq:bosact}) with an arbitrary given profile of magnetization along the string. To find such a minimum
one should solve a two-dimensional Laplace 
equation (obtained by variation of (\ref{eq:bosact}) with respect to $\phi$)
\begin{equation}
    \Delta\phi=0
 \label{eq:LL}
\end{equation}
with periodic boundary conditions
\begin{equation}
    \phi(z+i\beta)=\phi(z)
 \label{eq:bcperiod}
\end{equation}
and magnetization profile defined by
\begin{equation}
   \phi(x,\tau=0) = \frac{2}{Q}f(x), \;\mbox{when}\; 0<x<n.
 \label{eq:bcgen}
\end{equation}
Here as usual $Q=2/(c\pi R)$.
$f(x)$ is an arbitrary (real) function satisfying $f(0)=0$, $f(n)=n$ so that the total magnetization of the string is $cn$ and magnetization profile of the string is given by $\sigma^z(x)=c\partial_x f(x)$. It is easy to check that  the solution of Dirichlet problem (\ref{eq:LL}-\ref{eq:bcgen}) is given by
\begin{equation}
    \phi(z) =\frac{2\pi}{\beta Q} \R \int_0^n \frac{ds}{\pi i}\, \frac{f(s)}{\sinh\frac{\pi(s-z)}{\beta}}
    \left( \frac{\sinh\frac{\pi z}{\beta}\sinh\frac{\pi(z-n)}{\beta}}
            {\sinh\frac{\pi s}{\beta}\sinh\frac{\pi(s-n)}{\beta}} \right)^{\frac{1}{2}}.
 \label{eq:umgen}
\end{equation}
Indeed, it obviously satisfies Laplace equation and periodic boundary conditions. For real $z=x$ with $0<x<n$ the only real contribution to the integral can come from the half of the residue at $s=x$ which gives (\ref{eq:bcgen}).
In the following we restrict ourselves to the simplest case of a uniform magnetization $f(x)=x$ along the string. In this case we have
\begin{equation}
    \phi(x,\tau=0) = \frac{2x}{Q}, \;\mbox{when}\; 0<x<n.
 \label{eq:bcum}
\end{equation}
In the limit of zero temperature  $\beta\to \infty$ the solution of (\ref{eq:LL},\ref{eq:bcum})  is given by (one can either check it directly or obtain it from (\ref{eq:umgen}))
\begin{equation}
    \phi_0(z)=\frac{2}{Q} \R
    \left(z-\sqrt{z(z-n)}\right).
 \label{eq:phi0umR}
\end{equation}
We shift the position of the string for notational convenience to $-\tilde{n}<x<\tilde{n}$ with $\tilde{n}=n/2$ and obtain
\begin{equation}
    \phi_0(z)=\frac{2}{Q} \R
    \left(z-\sqrt{z^2-\tilde{n}^2)}\right).
 \label{eq:phi0um}
\end{equation}
One can recognize in (\ref{eq:phi0um}) the solution of a very well known problem of 
two-dimensional electrostatics. Namely, it is easy to see 
that $\phi_0$ is the potential created by metallic rod 
of the length $n$ inserted in the uniform 
electric field. We have for the  gradient of the field
$$
    \partial_x\phi_0-i\partial_\tau\phi_0 = \frac{2}{Q}
    \left(1-\frac{z}{\sqrt{z^{2}-\tilde{n}^2}}\right).
$$
The action of the optimal configuration $\phi_0$  is  given by a convergent integral 
$$ S[\phi_0] = \int d^{2}x\, \frac{1}{2}(\partial_\mu\phi_0)^2
=\frac{2}{Q^2}
\int d^{2}x\, \left|1-\frac{z}{\sqrt{z^{2}-\tilde{n}^2}}\right|^{2}
=\frac{2\pi \tilde{n}^2}{Q^2}= \frac{\pi n^2}{2Q^2}. $$
Finally, we obtain the leading contribution to the probability of uniformly ferromagnetic string $P_n\sim e^{-S(\phi_0)}$ as 
\begin{eqnarray}
    P_{n} &\sim & e^{-\alpha n^{2}},
  \label{eq:finprob} \\
    \alpha &=& \frac{\pi }{2Q^2}.
 \label{eq:finalpha}
\end{eqnarray}
We see that $\alpha$ for the string with uniform magnetization is much bigger\footnote{This is expected because we minimize the action in the smaller class of configurations.} than the one for a free string (with only total magnetization fixed) and corresponding probability decays much faster.

The solution (\ref{eq:phi0um}) is diverging near the ends of the string. This divergence is milder than the one for a free string ($\partial_\mu\phi_0 \sim r^{-1/2}$ instead of $r^{-1}$). Therefore, we expect bosonization to work even better in this case. Indeed, the estimate for $r_0$  ($\partial_\mu\phi_0(r_0) \sim \pi R$) gives $r_0\sim c^2 n$. Therefore, the short distance cutoff is defined as
$$
r_0 = \mbox{max}\left\{1,c^2 n\right\}.
$$
We see that for large strings $n\gg c^{-2}$ the action coming from the $r_0$ vicinities to the ends of the string is $\sim (\pi R r_0)^2 = c^2 n^2/Q^2\ll \alpha n^2$ and the small parameter justifying the use of bosonization is $c^2$ (compare to $1/\ln c^{-1}$ in the case of a free string).
Therefore, if one is trying to interpolate (illegally) our results to the case of the maximally ferromagnetic ($c\sim 1$) string one should use  (\ref{eq:finprob},\ref{eq:finalpha}) instead of (\ref{eq:PnrTshort}).

It is also easy to obtain from (\ref{eq:umgen}) ($f(x)=x$) the solution $\phi_0$ for for the optimal configuration in the limit of high temperature (very long strings). In this limit the main contribution to the action comes from $\phi_0(z)\approx 2x/Q$ when $0<x<n$ so that $S(\phi_0)\approx \frac{2\beta n}{Q^2}$ which gives
\begin{eqnarray}
    P_{n} &\sim & e^{-\gamma n},
  \label{eq:finTprob} \\
    \gamma &=& \frac{2\beta }{Q^2}.
 \label{eq:fingamma}
\end{eqnarray}
It is not a surprise that this result is identical to (\ref{eq:PnrTlong}) obtained for the free string. When string is very long the main contribution comes from the part of the solution which is linearly growing with distance  and is uniform in time direction. There is also a difference with the case of a free string in that the crossover from the Gaussian to the exponential behavior of $P_n$ occurs  at $n\sim \beta$ (not $n\sim \beta\ln c^{-1}$ as for the free string). The reason is the much milder divergence of the solution in the case of a uniform string.



\end{document}